\documentclass[reprint,prl,aps,superscriptaddress,amsmath,amssymb,twocolumn,showpacs]{revtex4-1}

\usepackage{graphicx} 
\usepackage{dcolumn} 
\usepackage{bm} 
\usepackage{amsmath}
\usepackage{braket}
\usepackage{natbib}
\usepackage{color}%
\usepackage{ulem}

\begin{document}


\title{Phonon Angular Momentum Induced by Temperature Gradient}




\author{Masato Hamada}
\affiliation{Department of Physics, Tokyo Institute of Technology, 2-12-1 Ookayama, Meguro-ku, Tokyo 152-8551, Japan}
\author{Emi Minamitani}
\affiliation{Department of Materials Engineering, University of Tokyo, 7-3-1 Hongo, Bunkyo-ku, Tokyo 113-8656, Japan}
\author{Motoaki Hirayama}
\affiliation{Department of Physics, Tokyo Institute of Technology, 2-12-1 Ookayama, Meguro-ku, Tokyo 152-8551, Japan}
\affiliation{%
TIES, Tokyo Institute of Technology, Ookayama, Meguro-ku, Tokyo 152-8551, Japan 
}%
\affiliation{Center for Emergent Mattar Science, RIKEN,
 2-1 Hirosawa, Wako, Saitama 351-0198, Japan}
\author{Shuichi Murakami}
\affiliation{Department of Physics, Tokyo Institute of Technology, 2-12-1 Ookayama, Meguro-ku, Tokyo 152-8551, Japan}
\affiliation{%
TIES, Tokyo Institute of Technology, Ookayama, Meguro-ku, Tokyo 152-8551, Japan 
}%



\date{\today}

\begin{abstract}
Phonon modes in crystals can have angular momenta in general. It nevertheless cancels in equilibrium when the time-reversal symmetry 
is preserved. In this paper we show that when a temperature gradient is applied and heat current flows in the crystal, the phonon distribution becomes off-equilibrium, 
and a finite angular momentum is generated by the heat current. This mechanism is analogous to the Edelstein effect in electronic systems. 
This effect requires crystals with sufficiently low crystallographic symmetries, such as polar or chiral crystal structures. Because of the positive charges of the nuclei, this phonon angular momentum induces magnetization.
In addition, when the crystal can freely rotate, this generated phonon angular momentum is converted to a rigid-body rotation of the crystal, due to the conservation of the total angular momentum. Furthermore, in metallic crystals, the phonon angular momentum 
 will be partially  converted into spin angular momentum of electrons. 
\end{abstract}
\maketitle




Conversions between the magnetization and the mechanical generation can be realized in various ways, such as  
Einstein-de Haas effect \cite{Einstein} and Barnett effect \cite{Barnett}. In the Einstein-de Haas effect, when the sample is magnetized by the external magnetic field, the sample rotates due to the conservation of the angular momentum. On the other hand, in the Barnett effect, a rotation of the sample induces magnetization. The key mechanism of  these effects is the spin-rotation coupling, which relates electronic spins with a 
mechanical rotation \cite{SRC02}. In addition, spin-rotation coupling
also enables mechanical generation of spin current in various systems
\cite{SHD01,SRC01,Masato01}.
In these effects, rotational motions of phonons in solids are important, and in this context
an phonon angular momentum is formulated theoretically \cite{PAM01,PAM02,PAM03}.

Here we focus on the phonon angular momentum introduced in \cite{PAM01}, which represents rotational motions of the 
nuclei within each phonon mode. 
In crystals with time-reversal symmetry, i.e. those without magnetic field or magnetization, the  phonon angular momentum of each mode is an odd function of  the wavevector 
$\bm{k}$
and the total angular momentum vanishes in equilibrium due to cancellation between phonons with the wavevector $\bm{k}$ and those with  $-\bm{k}$. 
Meanwhile, one can expect that this cancellation goes away by driving the system off the equilibrium, and nonzero phonon angular momentum is induced. 
In this letter, to theoretically show this scenario, we consider a crystal with a finite heat current, and show a nonzero total phonon angular momentum 
in the crystal due to its nonequilibrium phonon distribution.
The crystal symmetry should be sufficiently low to allow this effect, and in particular the inversion symmetry should be absent.
We calculate the phonon angular momentum generated by heat current for the wurtzite GaN as an example of polar systems and the Te (tellurium) and Se (selenium) as examples of chiral systems. For wurtzite GaN, we calculate the phonon properties by using the valence force field model and first-principle calculation, and for Te and Se, we calculate the phonon properties  by using the first-principle calculation.

The phonon angular momentum \cite{PAM01} is a part of the angular momentum of 
the microscopic local rotations of the nuclei around their equilibrium positions.
We begin with the eigenmode equation for phonons  
$D(\bm{k})\epsilon_{\sigma}(\bm{k}) = \omega_{\sigma}^2\epsilon_{\sigma}(\bm{k})$, 
where $\epsilon_{\sigma}(\bm{k})$ is a displacement polarization vector at the wave vector $\bm{k}$ with  a mode index $\sigma$, 
and $D$ is the dynamical matrix. Here, we set the normalization condition as $\epsilon_{\sigma}^{\dagger}(\bm{k})\epsilon_{\sigma}(\bm{k})=1$.
In equilibrium, 
the phonon angular momentum per unit volume  \cite{PAM01} is expressed as 
\begin{eqnarray}
J_{i}^{\rm{ph}} &=& \frac{1}{V} \sum_{\bm{k},\sigma} l_{\sigma,i}(\bm{k})\left[f_0(\omega_{\sigma}(\bm{k})) + \frac{1}{2}\right],\
i=x,y,z \label{PAM} \\
l_{\sigma,i}(\bm{k}) &=& \hbar \epsilon_{\sigma}^{\dagger}(\bm{k}) M_{i} \epsilon_{\sigma}(\bm{k}), \label{eachPAM}
\end{eqnarray}
where $f_0(\omega_{\sigma}(\bm{k})) = 1/(e^{\hbar\omega_{\sigma}(\bm{k})/k_{\rm{B}}T} -1)$ is the Bose distribution function, $\omega_{\sigma}(\bm{k})$ is the eigenfrequency of each mode, $T$ is the temperature, and $V$ denotes the sample volume. The matrix $M_{i}$ 
is the tensor product of the unit matrix and the generator of $SO(3)$ rotation for a unit cell with $N$ atoms given by $(M_{i})_{jk}=I_{N\times N}\otimes (-i)\varepsilon_{ijk}$ ($i,j,k=x,y,z$). $\bm{l}_{\sigma}(\bm{k})$ in Eq.~(\ref{eachPAM}) is the phonon angular momentum of a mode $\sigma$ at phonon wave vector $\bm{k}$, 
Because of the time-reversal symmetry of the system, it is an odd function of $\bm{k}$: $\bm{l}_{\sigma}(\bm{k}) =- \bm{l}_{\sigma}(-\bm{k})$, 
and their sum vanishes in equilibrium.




On the other hand, when the temperature gradient is nonzero, the phonon angular momentum becomes nonzero. 
Within the Boltzmann transport theory, the  distribution function deviates from the Bose distribution function $f_0$ as 
\begin{equation}
f_{\sigma,\bm{k}} = f_0(\omega_{\sigma}(\bm{k})) -\tau v_{\sigma,i}(\bm{k})\frac{\partial f_0}{\partial T} \frac{\partial T}{\partial x_i}, \label{Boltzmann}
\end{equation}
where 
$v_{\sigma,i}(\bm{k}) = \partial \omega_{\sigma}(\bm{k})/\partial k_i$ is the group velocity of each mode and $x_i$ 
is the $i$th component of the position. 
To justify the use of the Boltzmann transport theory, we assume here that the deviation of the system away from equilibrium is small. In order to satisfy this condition, we focus on the linear response regime where the heat current is infinitesimally small. We also assume that  the system relaxes towards the local thermal equilibrium quickly via nonlinear phonon-phonon interactions. 
As shown in Eq.~(\ref{Boltzmann}), the effect of nonlinear phonon-phonon interactions is represented by the phonon relaxation time $\tau$ based on the constant relaxation time approximation. 
The dependence of $\tau$ on the mode index $\sigma$ and the wavevector $\bm{k}$ does not alter our main conclusion and the constant relaxation time approximation is enough for a rough estimation.
By substituting Eq.~(\ref{Boltzmann}) into Eq.~(\ref{PAM}), the total phonon angular momentum per unit volume becomes 
\begin{equation}
J_{i}^{\rm{ph}} = -\frac{\tau}{V} \sum_{\bm{k},\sigma} l_{\sigma,i} v_{\sigma,j}\frac{\partial f_0(\omega_{\sigma}(\bm{k}))}{\partial T} \frac{\partial T}{\partial x_{j}}\equiv
 \alpha_{ij}\frac{\partial T}{\partial x_j} \label{HPAM}
\end{equation}
where $\alpha_{ij}$ denotes a response tensor. 
The generated phonon angular momentum is proportional to the temperature gradient.
This effect is caused by nonequilibrium phonon distribution, leading to an unbalance of phonon angular momentum, and therefore it is analogous to the Edelstein effect 
\cite{Ivchenko,Levitov,Aronov,edelstein1990spin,PhysRevB.67.033104,PhysRevLett.93.176601,orbitalEdelstein}
in electronic systems.

In order to realize a nonzero response tensor $\alpha_{ij}$, the crystal symmetry should be sufficiently low.
Necessary conditions for the crystallographic symmetry 
are shown in the Supplementary Material \cite{SM}. 
It is instructive to decompose the response tensor into symmetric and antisymmetric parts. 
The antisymmetric part of $\alpha_{ij}$ is essentially a polar vector $\alpha_k\equiv\epsilon_{ijk}\alpha_{ij}$
and therefore it survives only for polar crystals, such as ferroelectrics and polar metals. In this case, when we set the $z$ axis to
be along the  
polarization or the polar axis, $\alpha_{xy}=-\alpha_{yx}$ are the only nonzero elements of this tensor. Thus in any polar crystals, the temperature gradient and the generated angular momentum 
are perpendicular to each other, and they are both perpendicular to the polarization vector. 
On the other hand, the symmetric part of $\alpha_{ij}$ changes sign under inversion, and remains typically in chiral systems such as tellurium. 
In systems with very low symmetry, both the  antisymmetric and the symmetric parts 
become nonzero.

\begin{figure*}[thb]
\includegraphics[clip,width =15cm]{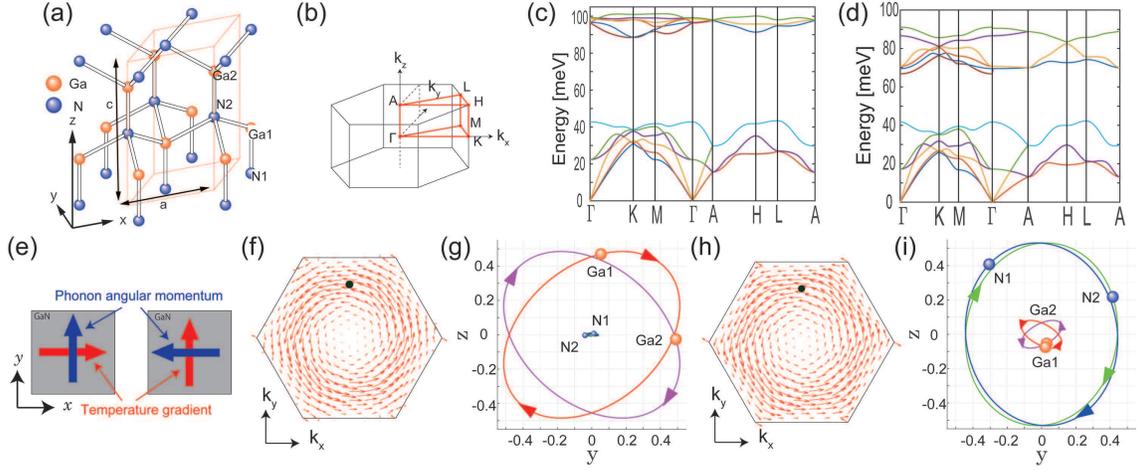}
\caption{\label{GaN} (color online) 
Crystal structure of GaN, and phonon angular momentum of GaN. 
(a) Crystal structure of wurtzite GaN.
The red line shows the unit cell. 
(b) First Brillouin zone of GaN. 
(c) Numerical results of phonon dispersion of wurtzite GaN using the valence force field model. 
(d) Numerical results of phonon dispersion of wurtzite GaN by first-principle calculation.
(e) Schematic illustration of the relation between  temperature gradient and the phonon angular momentum.
(f) Distribution of the phonon angular momentum $\bm{l}_{\sigma}(\bm{k})$ of the sixth lowest band on the plane $k_z=0$. 
(g) Trajectories of the four atoms in the unit cell for the phonon of the sixth band at $\frac{{\bf k}a}{2\pi}=(0,0.3849,0)$, 
which is indicated as a black dot in (f).
(h) Distribution of the phonon angular momentum $\bm{l}_{\sigma}(\bm{k})$ of the eleventh lowest band on the plane $k_z=0$. 
(i) Trajectories of the four atoms in the unit cell for the phonon of the eleventh band at $\frac{{\bf k}a}{2\pi}=(0,0.3849,0)$, 
which is indicated as a black dot in (h). 
(g) and (i) represent the normalized polarization vector $\varepsilon_{\sigma}({\bm k})$ with $\varepsilon^{\dagger}\varepsilon = 1$, and their axes $(x,y,z)$ are shown in a dimensionless unit.
}
\end{figure*}

As an example of polar systems, we discuss the wurtzite structure (space group: $P6_{3}mc$). The wurtzite structure 
has four atoms in the unit cell, as shown in Fig.~\ref{GaN}(a) for GaN.
When we take the polar axis to be along the $z$ axis, 
the nonzero elements of the response tensor $\alpha_{ij}$ are $\alpha_{xy}=-\alpha_{yx}$ from symmetry analysis, as  
shown in Fig.~\ref{GaN}(e). 
We first estimate the generated phonon angular momentum by heat current both by the valence force field model of Keating \cite{Parameter01} and by first-principle calculations. 
We describe the detalils of our first-principle calculation in the Supplementary Material~\cite{SM}.
The lattice structure and corresponding Brillouin zone are shown in Figs.~\ref{GaN}(a) and (b).
The details of valence force field model is summarized in the Supplementary Material \cite{SM}.
The result of phonon dispersion for wurtzite GaN by the valence force field model calculation is Fig.~\ref{GaN}(c).
Despite its simplicity, this model well describes the nature of phonon and phonon angular momentum can be evaluated (see Supplementary Material \cite{SM}).   
%
The band structure obtained by the first-principle calculation (Figs.~\ref{GaN}(d)) shows good agreement with previous works \cite{PhysRevLett.86.906,PhysRevB.61.6720,Parameter02}. Overall features of the band structure are similar to those obtained by the valence force field model except for the splitting of the longitudinal and transverse optical bands at the long wavelength limit. 
Examples of distributions of phonon angular momentum $\bm{l}_{\sigma}(\bm{k})$ are in
Figs.~\ref{GaN}(f) and (h), showing similarity with spin structure in Rashba systems. 
We show the trajectories of atoms in the sixth and eleventh lowest modes in Figs.~\ref{GaN}(g) and (i), respectively. By comparing Figs.~\ref{GaN}(g) and (i), the oscillation of nitrogen atoms in the eleventh modes is much larger than that of the gallium atoms, while the oscillation of gallium atoms in the sixth mode is larger.  
The response tensor is estimated as
$\alpha_{xy}\sim -10^{-7} \times [\tau/(1\text{s})] ~\rm{Jsm^{-2}K^{-1}}$ at $T = 300~{\rm K}$. 

\begin{figure*}
\includegraphics[width = 15cm]{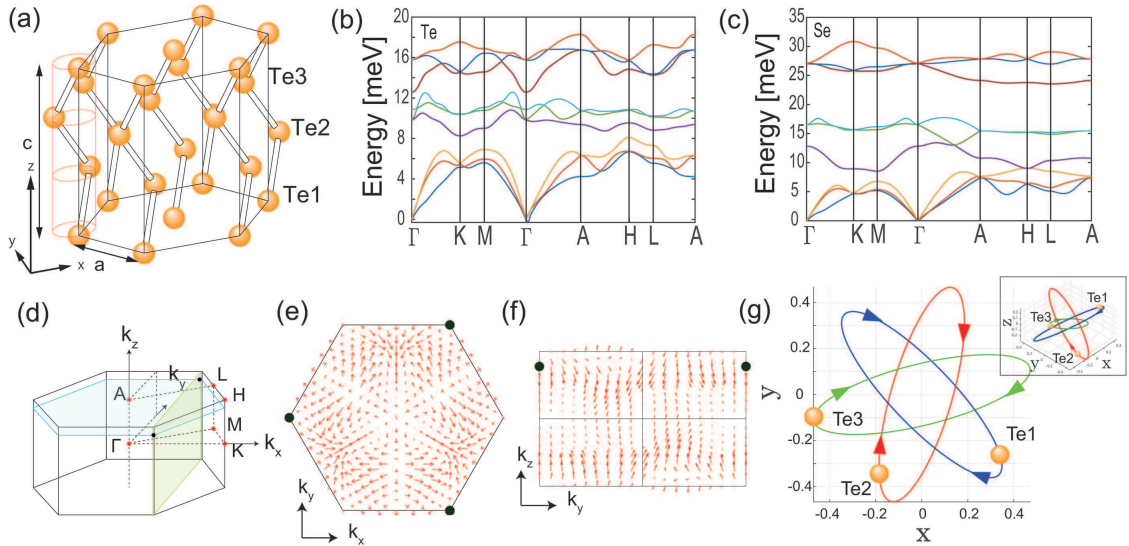}
\caption{\label{Te} (color online) Phonon angular momentum of Te and Se in the first-principle calculation.
(a) Crystal structure of Te. 
(b) Phonon dispersion of Te.
(c) Phonon dispersion of Se.
(d) First Brillouin zone of Te. 
(e), (f) Distribution of the phonon angular momentum $\bm{l}_{\sigma}(\bm{k})$ 
of the fourth lowest band. 
(e) and (f) show the results on the plane $\frac{k_za}{2\pi}= 0.2927$ and on the plane $\frac{k_xa}{2\pi} = \frac{1}{3}$, respectively. These planes correspond to the two cross sections in (d). 
(g) Trajectories of the four atoms in the unit cell for the phonon of the fourth lowest band at $\frac{{\bf k}a}{2\pi}=(\frac{1}{3},\frac{1}{\sqrt{3}},0.2927)$, 
which is indicated as the black dots in (d), (e), and (f). 
(g) represents the normalized polarization vector $\varepsilon_{\sigma}({\bm k})$ with $\varepsilon^{\dagger}\varepsilon = 1$, and its axis $(x,y,z)$ is shown in a dimensionless unit.
}
\end{figure*}

As other examples, we consider Te (tellurium) and Se (selenium) \cite{Hirayama-Te}. 
Te and Se have a helical crystal structure, as shown in Fig.~\ref{Te}(a).
The helical chains having three atoms in a unit cell 
form a triangular lattice. 
The space group is $P3_{1}21$ or $P3_{2}{21}$ ($D^{4}_{3}$ or $D^{6}_{3}$) 
corresponding to the right-handed or left-handed screw symmetry. 
They are semiconductors at ambient pressure.
Numerical results of the phonon dispersions of Te and Se by first-principle calculation are shown in 
Figs.~\ref{Te}(b) and (c), respectively.
The distributions of phonon angular momentum $\bm{l}_{\sigma}(\bm{k})$
on the two planes
in the Brillouin zone (Fig.~\ref{Te}(d)) are shown in Figs.~\ref{Te}(e), (f) 
for the fourth lowest band.
Because of the threefold screw symmetry around the $z$ axis, the
angular momentum on the $k_z$ axis is along the $z$ axis.  
Fig.~\ref{Te}(g) represents the trajectories of the 
three atoms in the unit cell for the fourth mode. 
Here, because of the threefold 
screw symmetry at the $\bm{k}$ point considered, the trajectories are related with each 
other by threefold rotation around the $z$ axis, and the angular momentum 
is along the $z$ axis by symmetry. 
The overall feature are similar between Se and Te, as shown in the Supplementary Material \cite{SM}. 
In Te and Se, from symmetry argument, the response tensor has nonzero elements 
$\alpha_{xx}=\alpha_{yy}$ and $\alpha_{zz}$, whose symmetry
 is identical with the electronic Edelstein effect 
in tellurium \cite{orbitalEdelstein}. The response tensor for  Te is estimated as
$
\alpha_{zz} \sim -10^{-7}\times [\tau_{\parallel}/(1s)] ~\rm{Jsm^{-2}K^{-1}}$ and 
$
\alpha_{xx} \sim 10^{-7}\times [\tau_{\perp}/(1s)]~\rm{Jsm^{-2}K^{-1}}$, and that
for Se is estimated as
$
\alpha_{zz} \sim -10^{-6}\times [\tau_{\parallel}/(1s)]~\rm{Jsm^{-2}K^{-1}}$ and 
$
\alpha_{xx} \sim -10^{-7}\times [\tau_{\perp}/(1s)]~\rm{Jsm^{-2}K^{-1}}$. 

\begin{figure}
\includegraphics[width=8cm]{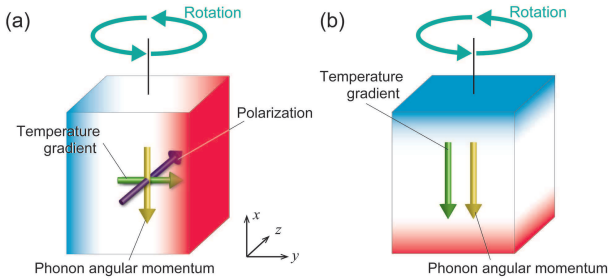}
\caption{\label{schematic}(color online) Schematic diagram of the generated rigid-body rotation due to the heat current.
(a) and (b) shows two typical cases: (a) for a polar crystal such as wurtzite GaN and (b) for a chiral crystal such as Te and Se.}
\end{figure}

Next we propose experiments to measure the phonon angular momentum generated by the heat current. 
The phonon angular momentum is a microscopic local rotation, and it cannot be 
measured directly.
To measure this, we consider a phonon version of the Einstein-de Haas effect. 
Suppose the crystal can rotate freely.
By conservation of angular momentum, when a heat current generates a phonon angular momentum $J^{\rm{ph}}$, a rigid-body rotation of the crystal also acquires an angular momentum which compensates the phonon angular momentum, $J^{\rm{rigid-body}} = -J^{\rm{ph}}$. 
This conservation holds when we take an average over a long period much longer than 
a typical time scale of the phonon motions, as discussed in detail in Supplemental Material.
For example, in polar crystals, when the heat current flows along the $y$-direction, the phonon angular momentum along the $x$-direction is generated, and 
it is converted to a rigid-body rotation, as shown in Fig.~\ref{schematic}(a).
Similarly, in tellurium, it is schematically shown in 
Fig.~\ref{schematic}(b), and the rotation direction will be opposite for right-handed and left-handed crystals.
 Next, we estimate the angular velocity $\omega$ of the rigid-body rotation in GaN as an example. We set the sample size to be $L\times L\times L$ and the phonon relaxation time to be $\tau \sim 10 ~\rm{ps}$ \cite{Parameter02}. The temperature difference over the sample size $L$ is denoted by $\Delta T$. 
The angular momentum of the rigid-body rotation is represented as $J^{\text{rigid-body}}L^3 = I\omega$, where
$I =  ML^2/6$ is the inertial moment of the sample with the total mass $M$. We estimate the  angular velocity of the rigid-body rotation as
\begin{equation}
\omega = -J^{ph}_{x}L^3/I \sim  \frac{\Delta T/(1\text{K})}{(L/(1\text{m}))^3} \times 10^{-21}~ {\rm s^{-1}}.
\end{equation}
Then, by setting the temperature difference to be $\Delta T = 10 ~\rm{K}$, an angular velocity of the rigid-body rotation $\omega$ is estimated as $10^{-8}~\rm{s}^{-1}$ when  
 $L=100~\rm{\mu m}$ and  $10^{-2}~\rm{s}^{-1}$ when $L=1~\rm{\mu m}$. They are sufficiently large for 
experimental measurement. The estimations for Te and Se are shown in the Supplementary Material \cite{SM}.

Because this phonon angular momentum means rotational motions of the nuclei having positive charges, 
it induces magnetization in itself. This can be estimated using a Born effective charge.
The magnetic moment $\bm{m}$ is related with the angular momentum $\bm{j}$ by 
$\bm{m} = \gamma \bm{j}$ with the gyromagnetic ratio $\gamma$. In the case of wurtzite GaN, the Born effective charge tensor $eZ^{*}_{\alpha \beta}$ is $eZ^{*}_{xx} = eZ^{*}_{yy} =  2.58e , eZ^{*}_{zz} = 2.71e$ from our {\it ab-initio} calculation. The gyromagnetic ratio tensors of the Ga and N atoms are given by $\gamma^{\rm{Ga}}_{\alpha\beta} = geZ^{*}_{\alpha\beta}/2m_{\rm Ga}$ and $\gamma^{\rm N}_{\alpha\beta} = -geZ^{*}_{\alpha\beta}/2m_{\rm N}$ with g-factor of GaN $g_{\parallel} = 1.951,g_{\perp} = 1.9483 $~\cite{wGaN-g}, where $m_{\rm Ga}$ and $m_{\rm N}$ are
the mass of the Ga atom and that of the N atom, respectively. We estimate the order of magnitude of the magnetization as
\begin{equation}
M_x \sim -\frac{\Delta T/(1\text{K})}{L/(1\text{m})} \times 10^{-11} ~{\rm Am^{-1}}.
\end{equation}
Therefore the magnetization $M_x$ of GaN induced by temperature gradient is estimated as $10^{-6} ~\rm{Am^{-1}}$ when $L=100 ~\mathrm{\mu m},~\Delta T = 10~\rm{K}$ and $10^{-4} ~\mathrm{Am^{-1}}$ when $L = 1~\mathrm{\mu m},~\Delta T = 10 ~\rm{K}$.
Although the order of magnitude of this magnetization is very small, it is expected to be observable experimentally.

In summary, we have theoretically predicted and estimated the phonon angular momentum generated by the heat current for wurtzite GaN, Te, and Se. This mechanism is analogous to the Edelstein effect in electronic systems.  We proposed experiments to measure the phonon angular momentum generated by the heat current. When the crystals can rotate freely, the phonon angular momentum generated by heat current is converted to a rigid-body rotation of the crystals due to the conservation of angular momentum. This rigid-body rotation is sufficiently large for experimental measurement when the size of sample is micro order. On the other hands, because of the  nuclei having positive charge, the phonon angular momentum generated by heat current induces magnetization.  
Moreover, in metals, the phonon angular momentum will be partially converted to electronic spin angular momentum via the spin-rotation coupling, which is similar to the spin-current generation 
proposed for the surface acoustic waves in solids 
\cite{SRC01}, and for the twiston modes in carbon nanotubes
\cite{Masato01}. These experimental proposals are expected to unveil 
properties of the phonon angular momentum.

\begin{acknowledgments}
This work was partly supported by a Grant-in-Aid for Scientific Research on Innovative Area, ``Nano Spin Conversion Science'' (Grant No.~26100006), by MEXT Elements Strategy Initiative to Form Core Research Center (TIES), and also by JSPS KAKENHI Grant Number JP17J10342.
\end{acknowledgments}



\begin{thebibliography}{23}%
\makeatletter
\providecommand \@ifxundefined [1]{%
 \@ifx{#1\undefined}
}%
\providecommand \@ifnum [1]{%
 \ifnum #1\expandafter \@firstoftwo
 \else \expandafter \@secondoftwo
 \fi
}%
\providecommand \@ifx [1]{%
 \ifx #1\expandafter \@firstoftwo
 \else \expandafter \@secondoftwo
 \fi
}%
\providecommand \natexlab [1]{#1}%
\providecommand \enquote  [1]{``#1''}%
\providecommand \bibnamefont  [1]{#1}%
\providecommand \bibfnamefont [1]{#1}%
\providecommand \citenamefont [1]{#1}%
\providecommand \href@noop [0]{\@secondoftwo}%
\providecommand \href [0]{\begingroup \@sanitize@url \@href}%
\providecommand \@href[1]{\@@startlink{#1}\@@href}%
\providecommand \@@href[1]{\endgroup#1\@@endlink}%
\providecommand \@sanitize@url [0]{\catcode `\\12\catcode `\$12\catcode
  `\&12\catcode `\#12\catcode `\^12\catcode `\_12\catcode `\%12\relax}%
\providecommand \@@startlink[1]{}%
\providecommand \@@endlink[0]{}%
\providecommand \url  [0]{\begingroup\@sanitize@url \@url }%
\providecommand \@url [1]{\endgroup\@href {#1}{\urlprefix }}%
\providecommand \urlprefix  [0]{URL }%
\providecommand \Eprint [0]{\href }%
\providecommand \doibase [0]{http://dx.doi.org/}%
\providecommand \selectlanguage [0]{\@gobble}%
\providecommand \bibinfo  [0]{\@secondoftwo}%
\providecommand \bibfield  [0]{\@secondoftwo}%
\providecommand \translation [1]{[#1]}%
\providecommand \BibitemOpen [0]{}%
\providecommand \bibitemStop [0]{}%
\providecommand \bibitemNoStop [0]{.\EOS\space}%
\providecommand \EOS [0]{\spacefactor3000\relax}%
\providecommand \BibitemShut  [1]{\csname bibitem#1\endcsname}%
\let\auto@bib@innerbib\@empty
\bibitem [{\citenamefont {Einstein}\ and\ \citenamefont
  {de~Haas}(1915)}]{Einstein}%
  \BibitemOpen
  \bibfield  {author} {\bibinfo {author} {\bibfnamefont {A.}~\bibnamefont
  {Einstein}}\ and\ \bibinfo {author} {\bibfnamefont {W.~J.}\ \bibnamefont
  {de~Haas}},\ }\href@noop {} {\bibfield {journal}{\bibinfo {journal} {Phys. Ges.}\ }\textbf {\bibinfo {volume} {17}},\ \bibinfo
  {pages} {152} (\bibinfo {year} {1915})}\BibitemShut {NoStop}%
\bibitem [{\citenamefont {Barnett}(1915)}]{Barnett}%
  \BibitemOpen
  \bibfield  {author} {\bibinfo {author} {\bibfnamefont {S.~J.}\ \bibnamefont
  {Barnett}},\ }
  {\bibfield  {journal}
  {\bibinfo  {journal} {Phys. Rev.}\ }\textbf {\bibinfo {volume} {6}},\
  \bibinfo {pages} {239} (\bibinfo {year} {1915})}\BibitemShut {NoStop}%
\bibitem [{\citenamefont {Matsuo}\ \emph {et~al.}(2011)\citenamefont {Matsuo},
  \citenamefont {Ieda}, \citenamefont {Saitoh},\ and\ \citenamefont
  {Maekawa}}]{SRC02}%
  \BibitemOpen
  \bibfield  {author} {\bibinfo {author} {\bibfnamefont {M.}~\bibnamefont
  {Matsuo}}, \bibinfo {author} {\bibfnamefont {J.}~\bibnamefont {Ieda}},
  \bibinfo {author} {\bibfnamefont {E.}~\bibnamefont {Saitoh}}, \ and\ \bibinfo
  {author} {\bibfnamefont {S.}~\bibnamefont {Maekawa}},\ }
  {\bibfield  {journal} {\bibinfo  {journal}
  {Phys. Rev. Lett.}\ }\textbf {\bibinfo {volume} {106}},\ \bibinfo {pages}
  {076601} (\bibinfo {year} {2011})}\BibitemShut {NoStop}%
\bibitem [{\citenamefont {Takahashi}\ \emph {et~al.}(2016)\citenamefont
  {Takahashi}, \citenamefont {Matsuo}, \citenamefont {Ono}, \citenamefont
  {Harii}, \citenamefont {Chudo}, \citenamefont {Okayasu}, \citenamefont
  {Ieda}, \citenamefont {Takahashi}, \citenamefont {Maekawa},\ and\
  \citenamefont {Saitoh}}]{SHD01}%
  \BibitemOpen
  \bibfield  {author} {\bibinfo {author} {\bibfnamefont {R.}~\bibnamefont
  {Takahashi}}, \bibinfo {author} {\bibfnamefont {M.}~\bibnamefont {Matsuo}},
  \bibinfo {author} {\bibfnamefont {M.}~\bibnamefont {Ono}}, \bibinfo {author}
  {\bibfnamefont {K.}~\bibnamefont {Harii}}, \bibinfo {author} {\bibfnamefont
  {H.}~\bibnamefont {Chudo}}, \bibinfo {author} {\bibfnamefont
  {S.}~\bibnamefont {Okayasu}}, \bibinfo {author} {\bibfnamefont
  {J.}~\bibnamefont {Ieda}}, \bibinfo {author} {\bibfnamefont {S.}~\bibnamefont
  {Takahashi}}, \bibinfo {author} {\bibfnamefont {S.}~\bibnamefont {Maekawa}},
  \ and\ \bibinfo {author} {\bibfnamefont {E.}~\bibnamefont {Saitoh}},\ }
  {\bibfield  {journal} {\bibinfo  {journal}
  {Nature Physics}\ }\textbf {\bibinfo {volume} {12}},\ \bibinfo {pages} {52}
  (\bibinfo {year} {2016})}\BibitemShut {NoStop}%
\bibitem [{\citenamefont {Matsuo}\ \emph {et~al.}(2013)\citenamefont {Matsuo},
  \citenamefont {Ieda}, \citenamefont {Harii}, \citenamefont {Saitoh},\ and\
  \citenamefont {Maekawa}}]{SRC01}%
  \BibitemOpen
  \bibfield  {author} {\bibinfo {author} {\bibfnamefont {M.}~\bibnamefont
  {Matsuo}}, \bibinfo {author} {\bibfnamefont {J.}~\bibnamefont {Ieda}},
  \bibinfo {author} {\bibfnamefont {K.}~\bibnamefont {Harii}}, \bibinfo
  {author} {\bibfnamefont {E.}~\bibnamefont {Saitoh}}, \ and\ \bibinfo {author}
  {\bibfnamefont {S.}~\bibnamefont {Maekawa}},\ }
  {\bibfield  {journal} {\bibinfo  {journal} {Phys.
  Rev. B}\ }\textbf {\bibinfo {volume} {87}},\ \bibinfo {pages} {180402}
  (\bibinfo {year} {2013})}\BibitemShut {NoStop}%
\bibitem [{\citenamefont {Hamada}\ \emph {et~al.}(2015)\citenamefont {Hamada},
  \citenamefont {Yokoyama},\ and\ \citenamefont {Murakami}}]{Masato01}%
  \BibitemOpen
  \bibfield  {author} {\bibinfo {author} {\bibfnamefont {M.}~\bibnamefont
  {Hamada}}, \bibinfo {author} {\bibfnamefont {T.}~\bibnamefont {Yokoyama}}, \
  and\ \bibinfo {author} {\bibfnamefont {S.}~\bibnamefont {Murakami}},\ }
  {\bibfield  {journal} {\bibinfo
  {journal} {Phys. Rev. B}\ }\textbf {\bibinfo {volume} {92}},\ \bibinfo
  {pages} {060409} (\bibinfo {year} {2015})}\BibitemShut {NoStop}%
\bibitem [{\citenamefont {Zhang}\ and\ \citenamefont {Niu}(2014)}]{PAM01}%
  \BibitemOpen
  \bibfield  {author} {\bibinfo {author} {\bibfnamefont {L.}~\bibnamefont
  {Zhang}}\ and\ \bibinfo {author} {\bibfnamefont {Q.}~\bibnamefont {Niu}},\
  }
  {\bibfield  {journal}
  {\bibinfo  {journal} {Phys. Rev. Lett.}\ }\textbf {\bibinfo {volume} {112}},\
  \bibinfo {pages} {085503} (\bibinfo {year} {2014})}\BibitemShut {NoStop}%
\bibitem [{\citenamefont {Garanin}\ and\ \citenamefont
  {Chudnovsky}(2015)}]{PAM02}%
  \BibitemOpen
  \bibfield  {author} {\bibinfo {author} {\bibfnamefont {D.~A.}\ \bibnamefont
  {Garanin}}\ and\ \bibinfo {author} {\bibfnamefont {E.~M.}\ \bibnamefont
  {Chudnovsky}},\ }
  {\bibfield
  {journal} {\bibinfo  {journal} {Phys. Rev. B}\ }\textbf {\bibinfo {volume}
  {92}},\ \bibinfo {pages} {024421} (\bibinfo {year} {2015})}\BibitemShut
  {NoStop}%
\bibitem [{\citenamefont {Zhang}\ and\ \citenamefont {Niu}(2015)}]{PAM03}%
  \BibitemOpen
  \bibfield  {author} {\bibinfo {author} {\bibfnamefont {L.}~\bibnamefont
  {Zhang}}\ and\ \bibinfo {author} {\bibfnamefont {Q.}~\bibnamefont {Niu}},\
  }
  {\bibfield  {journal}
  {\bibinfo  {journal} {Phys. Rev. Lett.}\ }\textbf {\bibinfo {volume} {115}},\
  \bibinfo {pages} {115502} (\bibinfo {year} {2015})}\BibitemShut {NoStop}%
\bibitem{Ivchenko}
E. L. Ivchenko and G. E. Pikus, JETP Lett. \textbf{27}, 604
(1978).
\bibitem{Levitov}
L. S. Levitov, Yu. V. Nazarov, and G. M. Eliashberg, Zh. \'{E}ksp.
Teor. Fiz. \textbf{88}, 229 (1985) [Sov. Phys. JETP \textbf{61}, 133 (1985)].
\bibitem{Aronov}
A. G. Aronov and Y. B. Lyanda-Geller, JETP Lett. \textbf{50},
431 (1989).
\bibitem [{\citenamefont {Edelstein}(1990)}]{edelstein1990spin}%
  \BibitemOpen
  \bibfield  {author} {\bibinfo {author} {\bibfnamefont {V.~M.}\ \bibnamefont
  {Edelstein}},\ }\href@noop {} {\bibfield  {journal} {\bibinfo  {journal}
  {Solid State Commun.}\ }\textbf {\bibinfo {volume} {73}},\ \bibinfo
  {pages} {233} (\bibinfo {year} {1990})}\BibitemShut {NoStop}%
\bibitem [{\citenamefont {Inoue}\ \emph {et~al.}(2003)\citenamefont {Inoue},
  \citenamefont {Bauer},\ and\ \citenamefont {Molenkamp}}]{PhysRevB.67.033104}%
  \BibitemOpen
  \bibfield  {author} {\bibinfo {author} {\bibfnamefont {J.-i.}\ \bibnamefont
  {Inoue}}, \bibinfo {author} {\bibfnamefont {G.~E.~W.}\ \bibnamefont {Bauer}},
  \ and\ \bibinfo {author} {\bibfnamefont {L.~W.}\ \bibnamefont {Molenkamp}},\
  }
  {\bibfield  {journal} {\bibinfo
  {journal} {Phys. Rev. B}\ }\textbf {\bibinfo {volume} {67}},\ \bibinfo
  {pages} {033104} (\bibinfo {year} {2003})}\BibitemShut {NoStop}%
\bibitem [{\citenamefont {Kato}\ \emph {et~al.}(2004)\citenamefont {Kato},
  \citenamefont {Myers}, \citenamefont {Gossard},\ and\ \citenamefont
  {Awschalom}}]{PhysRevLett.93.176601}%
  \BibitemOpen
  \bibfield  {author} {\bibinfo {author} {\bibfnamefont {Y.~K.}\ \bibnamefont
  {Kato}}, \bibinfo {author} {\bibfnamefont {R.~C.}\ \bibnamefont {Myers}},
  \bibinfo {author} {\bibfnamefont {A.~C.}\ \bibnamefont {Gossard}}, \ and\
  \bibinfo {author} {\bibfnamefont {D.~D.}\ \bibnamefont {Awschalom}},\ }
  {\bibfield  {journal} {\bibinfo
  {journal} {Phys. Rev. Lett.}\ }\textbf {\bibinfo {volume} {93}},\ \bibinfo
  {pages} {176601} (\bibinfo {year} {2004})}\BibitemShut {NoStop}%
\bibitem [{\citenamefont {Yoda}\ \emph {et~al.}(2015)\citenamefont {Yoda},
  \citenamefont {Yokoyama},\ and\ \citenamefont {Murakami}}]{orbitalEdelstein}%
  \BibitemOpen
  \bibfield  {author} {\bibinfo {author} {\bibfnamefont {T.}~\bibnamefont
  {Yoda}}, \bibinfo {author} {\bibfnamefont {T.}~\bibnamefont {Yokoyama}}, \
  and\ \bibinfo {author} {\bibfnamefont {S.}~\bibnamefont {Murakami}},\
  }\href@noop {} {\bibfield  {journal} {\bibinfo  {journal} {Sci. Rep.}\
  }\textbf {\bibinfo {volume} {5}},\ \bibinfo {pages} {12024} (\bibinfo {year}
  {2015})}\BibitemShut {NoStop}%
\bibitem [{SM()}]{SM}%
  \BibitemOpen
  \href@noop {} {}\bibinfo {note} {See Supplemental Material, which includes Refs.~\cite{1962series,RevModPhys.73.515,QE,GBRV,PhysRevB.13.5188,Coorper,Springer,doi:10.1063/1.4938203}, for detailed discussion and details of calculations }\BibitemShut {Stop}%
\bibitem [{\citenamefont {Birss}(1962)}]{1962series}%
  \BibitemOpen
  \bibfield  {author} {\bibinfo {author} {\bibfnamefont {R.~R.}\ \bibnamefont
  {Birss}},\ }
  {
  \textit{\bibinfo {title} {Symmetry and Magnetism, Series of Monographs on Selected
  Topics in Solid State Physics}}} ,\ edited by\ \bibinfo {editor}
  {\bibfnamefont {E.~P.}\ \bibnamefont {Wohlfarth}},\ \bibinfo {series} {Series
  of Monographs on Selected Topics in Solid State Physics}, Vol.~\bibinfo
  {volume} {3}\ (\bibinfo  {publisher} {Elsevier North-Holland},\ \bibinfo
  {year} {1962})\BibitemShut {NoStop}%
\bibitem [{\citenamefont {Baroni}\ \emph {et~al.}(2001)\citenamefont {Baroni},
  \citenamefont {de~Gironcoli}, \citenamefont {Dal~Corso},\ and\ \citenamefont
  {Giannozzi}}]{RevModPhys.73.515}%
  \BibitemOpen
  \bibfield  {author} {\bibinfo {author} {\bibfnamefont {S.}~\bibnamefont
  {Baroni}}, \bibinfo {author} {\bibfnamefont {S.}~\bibnamefont
  {de~Gironcoli}}, \bibinfo {author} {\bibfnamefont {A.}~\bibnamefont
  {Dal~Corso}}, \ and\ \bibinfo {author} {\bibfnamefont {P.}~\bibnamefont
  {Giannozzi}},\ }
  {\bibfield
  {journal} {\bibinfo  {journal} {Rev. Mod. Phys.}\ }\textbf {\bibinfo {volume}
  {73}},\ \bibinfo {pages} {515} (\bibinfo {year} {2001})}\BibitemShut
  {NoStop}%
\bibitem [{\citenamefont {Giannozzi}\ \textit {et~al.}(2009)\citenamefont
  {Giannozzi}, \citenamefont {Baroni}, \citenamefont {Bonini}, \citenamefont
  {Calandra}, \citenamefont {Car}, \citenamefont {Cavazzoni}, \citenamefont
  {Ceresoli}, \citenamefont {Chiarotti}, \citenamefont {Cococcioni},
  \citenamefont {Dabo}, \citenamefont {Corso}, \citenamefont {de~Gironcoli},
  \citenamefont {Fabris}, \citenamefont {Fratesi}, \citenamefont {Gebauer},
  \citenamefont {Gerstmann}, \citenamefont {Gougoussis}, \citenamefont
  {Kokalj}, \citenamefont {Lazzeri}, \citenamefont {Martin-Samos},
  \citenamefont {Marzari}, \citenamefont {Mauri}, \citenamefont {Mazzarello},
  \citenamefont {Paolini}, \citenamefont {Pasquarello}, \citenamefont
  {Paulatto}, \citenamefont {Sbraccia}, \citenamefont {Scandolo}, \citenamefont
  {Sclauzero}, \citenamefont {Seitsonen}, \citenamefont {Smogunov},
  \citenamefont {Umari},\ and\ \citenamefont {Wentzcovitch}}]{QE}%
  \BibitemOpen
  \bibfield  {author} {\bibinfo {author} {\bibfnamefont {P.}~\bibnamefont
  {Giannozzi}}, \bibinfo {author} {\bibfnamefont {S.}~\bibnamefont {Baroni}},
  \bibinfo {author} {\bibfnamefont {N.}~\bibnamefont {Bonini}}, \bibinfo
  {author} {\bibfnamefont {M.}~\bibnamefont {Calandra}}, \bibinfo {author}
  {\bibfnamefont {R.}~\bibnamefont {Car}}, \bibinfo {author} {\bibfnamefont
  {C.}~\bibnamefont {Cavazzoni}}, \bibinfo {author} {\bibfnamefont
  {D.}~\bibnamefont {Ceresoli}}, \bibinfo {author} {\bibfnamefont {G.~L.}\
  \bibnamefont {Chiarotti}}, \bibinfo {author} {\bibfnamefont {M.}~\bibnamefont
  {Cococcioni}}, \bibinfo {author} {\bibfnamefont {I.}~\bibnamefont {Dabo}},
  \bibinfo {author} {\bibfnamefont {A.~D.}\ \bibnamefont {Corso}}, \bibinfo
  {author} {\bibfnamefont {S.}~\bibnamefont {de~Gironcoli}}, \bibinfo {author}
  {\bibfnamefont {S.}~\bibnamefont {Fabris}}, \bibinfo {author} {\bibfnamefont
  {G.}~\bibnamefont {Fratesi}}, \bibinfo {author} {\bibfnamefont
  {R.}~\bibnamefont {Gebauer}}, \bibinfo {author} {\bibfnamefont
  {U.}~\bibnamefont {Gerstmann}}, \bibinfo {author} {\bibfnamefont
  {C.}~\bibnamefont {Gougoussis}}, \bibinfo {author} {\bibfnamefont
  {A.}~\bibnamefont {Kokalj}}, \bibinfo {author} {\bibfnamefont
  {M.}~\bibnamefont {Lazzeri}}, \bibinfo {author} {\bibfnamefont
  {L.}~\bibnamefont {Martin-Samos}}, \bibinfo {author} {\bibfnamefont
  {N.}~\bibnamefont {Marzari}}, \bibinfo {author} {\bibfnamefont
  {F.}~\bibnamefont {Mauri}}, \bibinfo {author} {\bibfnamefont
  {R.}~\bibnamefont {Mazzarello}}, \bibinfo {author} {\bibfnamefont
  {S.}~\bibnamefont {Paolini}}, \bibinfo {author} {\bibfnamefont
  {A.}~\bibnamefont {Pasquarello}}, \bibinfo {author} {\bibfnamefont
  {L.}~\bibnamefont {Paulatto}}, \bibinfo {author} {\bibfnamefont
  {C.}~\bibnamefont {Sbraccia}}, \bibinfo {author} {\bibfnamefont
  {S.}~\bibnamefont {Scandolo}}, \bibinfo {author} {\bibfnamefont
  {G.}~\bibnamefont {Sclauzero}}, \bibinfo {author} {\bibfnamefont {A.~P.}\
  \bibnamefont {Seitsonen}}, \bibinfo {author} {\bibfnamefont {A.}~\bibnamefont
  {Smogunov}}, \bibinfo {author} {\bibfnamefont {P.}~\bibnamefont {Umari}}, \
  and\ \bibinfo {author} {\bibfnamefont {R.~M.}\ \bibnamefont {Wentzcovitch}},\
  }
  {\bibfield
  {journal} {\bibinfo  {journal} {Journal of Physics: Condensed Matter}\
  }\textbf {\bibinfo {volume} {21}},\ \bibinfo {pages} {395502} (\bibinfo
  {year} {2009})}\BibitemShut {NoStop}%
\bibitem [{\citenamefont {Garrity}\ \emph {et~al.}(2014)\citenamefont
  {Garrity}, \citenamefont {Bennett}, \citenamefont {Rabe},\ and\ \citenamefont
  {Vanderbilt}}]{GBRV}%
  \BibitemOpen
  \bibfield  {author} {\bibinfo {author} {\bibfnamefont {K.~F.}\ \bibnamefont
  {Garrity}}, \bibinfo {author} {\bibfnamefont {J.~W.}\ \bibnamefont
  {Bennett}}, \bibinfo {author} {\bibfnamefont {K.~M.}\ \bibnamefont {Rabe}}, \
  and\ \bibinfo {author} {\bibfnamefont {D.}~\bibnamefont {Vanderbilt}},\
  }
  {\bibfield  {journal} {\bibinfo  {journal} {Computational Materials Science}\
  }\textbf {\bibinfo {volume} {81}},\ \bibinfo {pages} {446 } (\bibinfo {year}
  {2014})}\BibitemShut {NoStop}%
\bibitem [{\citenamefont {Monkhorst}\ and\ \citenamefont
  {Pack}(1976)}]{PhysRevB.13.5188}%
  \BibitemOpen
  \bibfield  {author} {\bibinfo {author} {\bibfnamefont {H.~J.}\ \bibnamefont
  {Monkhorst}}\ and\ \bibinfo {author} {\bibfnamefont {J.~D.}\ \bibnamefont
  {Pack}},\ }
  {\bibfield  {journal}
  {\bibinfo  {journal} {Phys. Rev. B}\ }\textbf {\bibinfo {volume} {13}},\
  \bibinfo {pages} {5188} (\bibinfo {year} {1976})}\BibitemShut {NoStop}%
\bibitem [{\citenamefont {Cooper}(1969)}]{Coorper}%
  \BibitemOpen
  \bibinfo {editor} {\bibfnamefont {C.}~\bibnamefont {Cooper}, \bibfnamefont
  {W}},\ ed.,\ \href@noop {} {
  \textit{\bibinfo {title} {The Physics of Selenium
  and Tellurium}}}\ (\bibinfo  {publisher} {Pergamon},\ \bibinfo {year}
  {1969})\BibitemShut {NoStop}%
\bibitem [{\citenamefont {Gerlach}\ and\ \citenamefont
  {Grosse}(1979)}]{Springer}%
  \BibitemOpen
  \bibinfo {editor} {\bibfnamefont {E.}~\bibnamefont {Gerlach}}\ and\ \bibinfo
  {editor} {\bibfnamefont {P.}~\bibnamefont {Grosse}},\ eds.,\ \href@noop {}
  {
  \textit{\bibinfo {title} {The Physics of Selenium and Tellurium}}},\ \bibinfo
  {series} {Springer Series in Solid-State Sciences}, Vol.~\bibinfo {volume}
  {13}\ (\bibinfo  {publisher} {Springer-Verlag Berlin Heidelberg},\ \bibinfo
  {year} {1979})\BibitemShut {NoStop}%
\bibitem [{\citenamefont {Peng}\ \emph {et~al.}(2015)\citenamefont {Peng},
  \citenamefont {Kioussis},\ and\ \citenamefont
  {Stewart}}]{doi:10.1063/1.4938203}%
  \BibitemOpen
  \bibfield  {author} {\bibinfo {author} {\bibfnamefont {H.}~\bibnamefont
  {Peng}}, \bibinfo {author} {\bibfnamefont {N.}~\bibnamefont {Kioussis}}, \
  and\ \bibinfo {author} {\bibfnamefont {D.~A.}\ \bibnamefont {Stewart}},\
  }
  {\bibfield  {journal} {\bibinfo
  {journal} {Applied Physics Letters}\ }\textbf {\bibinfo {volume} {107}},\
  \bibinfo {pages} {251904} (\bibinfo {year} {2015})},
  \BibitemShut {NoStop}%
\bibitem [{\citenamefont {Camacho}\ and\ \citenamefont
  {Niquet}(2010)}]{Parameter01}%
  \BibitemOpen
  \bibfield  {author} {\bibinfo {author} {\bibfnamefont {D.}~\bibnamefont
  {Camacho}}\ and\ \bibinfo {author} {\bibfnamefont {Y.}~\bibnamefont
  {Niquet}},\ }
  {\bibfield  {journal} {\bibinfo  {journal} {Physica E: Low-dimensional
  Systems and Nanostructures}\ }\textbf {\bibinfo {volume} {42}},\ \bibinfo
  {pages} {1361 } (\bibinfo {year} {2010})}\BibitemShut {NoStop}%
\bibitem [{\citenamefont {Ruf}\ \emph {et~al.}(2001)\citenamefont {Ruf},
  \citenamefont {Serrano}, \citenamefont {Cardona}, \citenamefont {Pavone},
  \citenamefont {Pabst}, \citenamefont {Krisch}, \citenamefont {D'Astuto},
  \citenamefont {Suski}, \citenamefont {Grzegory},\ and\ \citenamefont
  {Leszczynski}}]{PhysRevLett.86.906}%
  \BibitemOpen
  \bibfield  {author} {\bibinfo {author} {\bibfnamefont {T.}~\bibnamefont
  {Ruf}}, \bibinfo {author} {\bibfnamefont {J.}~\bibnamefont {Serrano}},
  \bibinfo {author} {\bibfnamefont {M.}~\bibnamefont {Cardona}}, \bibinfo
  {author} {\bibfnamefont {P.}~\bibnamefont {Pavone}}, \bibinfo {author}
  {\bibfnamefont {M.}~\bibnamefont {Pabst}}, \bibinfo {author} {\bibfnamefont
  {M.}~\bibnamefont {Krisch}}, \bibinfo {author} {\bibfnamefont
  {M.}~\bibnamefont {D'Astuto}}, \bibinfo {author} {\bibfnamefont
  {T.}~\bibnamefont {Suski}}, \bibinfo {author} {\bibfnamefont
  {I.}~\bibnamefont {Grzegory}}, \ and\ \bibinfo {author} {\bibfnamefont
  {M.}~\bibnamefont {Leszczynski}},\ }
  {\bibfield  {journal} {\bibinfo  {journal} {Phys.
  Rev. Lett.}\ }\textbf {\bibinfo {volume} {86}},\ \bibinfo {pages} {906}
  (\bibinfo {year} {2001})}\BibitemShut {NoStop}%
\bibitem [{\citenamefont {Bungaro}\ \emph {et~al.}(2000)\citenamefont
  {Bungaro}, \citenamefont {Rapcewicz},\ and\ \citenamefont
  {Bernholc}}]{PhysRevB.61.6720}%
  \BibitemOpen
  \bibfield  {author} {\bibinfo {author} {\bibfnamefont {C.}~\bibnamefont
  {Bungaro}}, \bibinfo {author} {\bibfnamefont {K.}~\bibnamefont {Rapcewicz}},
  \ and\ \bibinfo {author} {\bibfnamefont {J.}~\bibnamefont {Bernholc}},\
  }
  {\bibfield  {journal} {\bibinfo
  {journal} {Phys. Rev. B}\ }\textbf {\bibinfo {volume} {61}},\ \bibinfo
  {pages} {6720} (\bibinfo {year} {2000})}\BibitemShut {NoStop}%
\bibitem [{\citenamefont {Wu}\ \emph {et~al.}(2016)\citenamefont {Wu},
  \citenamefont {Lee}, \citenamefont {Varshney}, \citenamefont {Wohlwend},
  \citenamefont {Roy},\ and\ \citenamefont {Luo}}]{Parameter02}%
  \BibitemOpen
  \bibfield  {author} {\bibinfo {author} {\bibfnamefont {X.}~\bibnamefont
  {Wu}}, \bibinfo {author} {\bibfnamefont {J.}~\bibnamefont {Lee}}, \bibinfo
  {author} {\bibfnamefont {V.}~\bibnamefont {Varshney}}, \bibinfo {author}
  {\bibfnamefont {J.~L.}\ \bibnamefont {Wohlwend}}, \bibinfo {author}
  {\bibfnamefont {A.~K.}\ \bibnamefont {Roy}}, \ and\ \bibinfo {author}
  {\bibfnamefont {T.}~\bibnamefont {Luo}},\ }\href@noop {} {\bibfield
  {journal} {\bibinfo  {journal} {Scientific Reports}\ }\textbf {\bibinfo
  {volume} {6}},\ \bibinfo {pages} {22504} (\bibinfo {year}
  {2016})}\BibitemShut {NoStop}%
\bibitem [{\citenamefont {Hirayama}\ \emph {et~al.}(2015)\citenamefont
  {Hirayama}, \citenamefont {Okugawa}, \citenamefont {Ishibashi}, \citenamefont
  {Murakami},\ and\ \citenamefont {Miyake}}]{Hirayama-Te}%
  \BibitemOpen
  \bibfield  {author} {\bibinfo {author} {\bibfnamefont {M.}~\bibnamefont
  {Hirayama}}, \bibinfo {author} {\bibfnamefont {R.}~\bibnamefont {Okugawa}},
  \bibinfo {author} {\bibfnamefont {S.}~\bibnamefont {Ishibashi}}, \bibinfo
  {author} {\bibfnamefont {S.}~\bibnamefont {Murakami}}, \ and\ \bibinfo
  {author} {\bibfnamefont {T.}~\bibnamefont {Miyake}},\ }
  {\bibfield  {journal} {\bibinfo  {journal}
  {Phys. Rev. Lett.}\ }\textbf {\bibinfo {volume} {114}},\ \bibinfo {pages}
  {206401} (\bibinfo {year} {2015})}\BibitemShut {NoStop}%
\bibitem [{\citenamefont {Carlos}\ \emph {et~al.}(1993)\citenamefont {Carlos},
  \citenamefont {Freitas}, \citenamefont {Khan}, \citenamefont {Olson},\ and\
  \citenamefont {Kuznia}}]{wGaN-g}%
  \BibitemOpen
  \bibfield  {author} {\bibinfo {author} {\bibfnamefont {W.~E.}\ \bibnamefont
  {Carlos}}, \bibinfo {author} {\bibfnamefont {J.~A.}\ \bibnamefont {Freitas}},
  \bibinfo {author} {\bibfnamefont {M.~A.}\ \bibnamefont {Khan}}, \bibinfo
  {author} {\bibfnamefont {D.~T.}\ \bibnamefont {Olson}}, \ and\ \bibinfo
  {author} {\bibfnamefont {J.~N.}\ \bibnamefont {Kuznia}},\ }
  {\bibfield  {journal}
  {\bibinfo  {journal} {Phys.
  Rev. B}\ }\textbf {\bibinfo {volume} {48}},\ \bibinfo {pages} {17878}
  (\bibinfo {year} {1993})}\BibitemShut {NoStop}%
\end{thebibliography}
%








\end{document}



\title{Phonon Angular Momentum Induced by Temperature Gradient : Supplemental Material}

\author{Masato Hamada}
\affiliation{Department of Physics, Tokyo Institute of Technology, 2-12-1 Ookayama, Meguro-ku, Tokyo 152-8551, Japan}
\author{Emi Minamitani}
\affiliation{Department of Materials Engineering, University of Tokyo, 7-3-1 Hongo, Bunkyo-ku, Tokyo 113-8656, Japan}
\author{Motoaki Hirayama}
\affiliation{Department of Physics, Tokyo Institute of Technology, 2-12-1 Ookayama, Meguro-ku, Tokyo 152-8551, Japan}
\affiliation{%
TIES, Tokyo Institute of Technology, Ookayama, Meguro-ku, Tokyo 152-8551, Japan 
}%
\affiliation{Center for Emergent Mattar Science, RIKEN,
2-1 Hirosawa, Wako, Saitama 351-0198, Japan}
\author{Shuichi Murakami}
\affiliation{Department of Physics, Tokyo Institute of Technology, 2-12-1 Ookayama, Meguro-ku, Tokyo 152-8551, Japan}
\affiliation{%
TIES, Tokyo Institute of Technology, Ookayama, Meguro-ku, Tokyo 152-8551, Japan 
}%


\maketitle

\section{Symmetry constraint for phonon Edelstein effect}
In the main text we discussed the angular momentum of phonons $\bm{J}^{\rm{ph}}$ induced
by the heat current, i.e. phonon Edelstein effect. Its response tensor $\alpha_{ij}$ 
is defined by
\begin{equation}
J_{i}^{\rm{ph}} =
 \alpha_{ij}\frac{\partial T}{\partial x_j}. \label{HPAM}
\end{equation}
In this section we discuss the constraints on the tensor $\alpha_{ij}$
by crystallographic symmetry. 
Because it is a response at $\bm{q}=0$, the nonzero elements 
of the tensor are determined only by the point-group symmetry, and 
not by the space-group symmetry. 
By noting that $\alpha_{ij}$ is an axial tensor, one
can easily identify nonzero elements of the tensor. 
First, among 32 point groups, the tensor vanishes in the $11$ point groups having inversion symmetry.
Among the remaining $21$ point groups without inversion symmetry, the axial tensor $\alpha_{ij}$ has  nonzero elements in $18$ point groups: $O$, $T$, $D_4$, $D_{2d}$, $C_{4v}$, $S_4$, $C_4$, $D_2$, $C_{2v}$, $D_{6}$, $C_{6v}$, $C_{6}$, $D_{3}$, $C_{3v}$, $C_{3}$, $C_{1h}$, $C_2$, and $C_1$. For example, we show the axial tensor $\alpha_{ij}$ in systems with the point group $C_{6v}$. The generators of the point group $C_{6v}$ are represented as 
\begin{equation}
C_{6} = 
\begin{pmatrix}
\frac{1}{2} & -\frac{\sqrt{3}}{2} & 0 \\
\frac{\sqrt{3}}{2} & \frac{1}{2} & 0 \\
0 & 0 & 1
\end{pmatrix}
,\ \sigma_{x} = 
\begin{pmatrix}
-1 & 0 & 0 \\
0 & 1 & 0 \\
0 & 0 & 1 
\end{pmatrix}.
\end{equation}
Because the axial tensor $\alpha$ should satisfy $C_{6}\alpha C_{6}^{-1} = \alpha$ and $-\sigma_x \alpha \sigma_x^{-1} = \alpha$, where $\alpha$ is the matrix consisting of
the elements $\alpha_{ij}$, the tensor is represented as
\begin{equation}
\alpha = 
\begin{pmatrix}
0 & \alpha_{xy} & 0 \\
-\alpha_{xy} & 0 & 0 \\
0 & 0 & 0 
\end{pmatrix}.
\end{equation}
In fact, a full list of the form of the axial tensor for the $18$ point groups is
available from \cite{1962series}, as follows:
\begin{eqnarray}
&\alpha =&
\begin{pmatrix}
\alpha_{xx} & 0 & 0 \\
0 & \alpha_{xx} & 0 \\
0 & 0 & \alpha_{xx} 
\end{pmatrix}: \ \ O,\ T,
\label{OT}
\\&
\alpha =&
\begin{pmatrix}
\alpha_{xx} & 0 & 0 \\
0 & \alpha_{xx} & 0 \\
0 & 0 & \alpha_{zz} 
\end{pmatrix}: \ \ D_4,\ D_6,\ D_3,
\label{TeSe}
\\&
\alpha =&
\begin{pmatrix}
\alpha_{xx} & \alpha_{xy} & 0 \\
-\alpha_{xy} & \alpha_{xx} & 0 \\
0 & 0 & \alpha_{zz} 
\end{pmatrix}: \ \ C_4,\ C_{6},\ C_{3},
\\&
\alpha =&
\begin{pmatrix}
\alpha_{xx} & \alpha_{xy} & 0 \\
\alpha_{xy} & -\alpha_{xx} & 0 \\
0 & 0 & 0 
\end{pmatrix}: \ \  S_4,
\\&
\alpha =&
\begin{pmatrix}
0 & \alpha_{xy} & 0 \\
-\alpha_{xy} & 0 & 0 \\
0 & 0 & 0 
\end{pmatrix}:\ \ C_{4v},\ C_{6v},\ C_{3v},
\label{GaN}
\\&
\alpha =&
\begin{pmatrix}
\alpha_{xx} & 0 & 0 \\
0 & -\alpha_{xx} & 0 \\
0 & 0 & 0 
\end{pmatrix}:\ \ D_{2d},
\\&
\alpha =&
\begin{pmatrix}
\alpha_{xx} & 0 & 0 \\
0 & \alpha_{yy} & 0 \\
0 & 0 & \alpha_{zz} 
\end{pmatrix}:\ \ D_{2},
\\&
\alpha =&
\begin{pmatrix}
0 & \alpha_{xy} & 0 \\
\alpha_{yx} & 0 & 0 \\
0 & 0 & 0 
\end{pmatrix}:\ \ C_{2v},
\\&
\alpha =&
\begin{pmatrix}
0 & 0 & \alpha_{xz} \\
0 & 0 & \alpha_{yz} \\
\alpha_{zx} & \alpha_{zy} & 0 
\end{pmatrix}:\ \  C_{1h},
\\&
\alpha =&
\begin{pmatrix}
\alpha_{xx} & \alpha_{xy} & 0 \\
\alpha_{yx} & \alpha_{yy} & 0 \\
0 & 0 & \alpha_{zz} 
\end{pmatrix}:\ \ C_2,
\\&
\alpha =&
\begin{pmatrix}
\alpha_{xx} & \alpha_{xy} & \alpha_{xz} \\
\alpha_{yx} & \alpha_{yy} & \alpha_{yz} \\
\alpha_{zx} & \alpha_{zy} & \alpha_{zz} 
\end{pmatrix}:\ \ C_1.
\label{C1}
\end{eqnarray}
Lastly we note that the mechanism of this effect is by nonequilibrium phonon distribution inducing an unbalance of phonon angular momentum, and it is analogous to the Edelstein effect 
\cite{Ivchenko,Levitov,Aronov,edelstein1990spin,PhysRevB.67.033104,PhysRevLett.93.176601,orbitalEdelstein}
in electronic systems. Therefore, symmetry constraints described 
in Eqs.~(\ref{OT})-(\ref{C1}) 
 are the same 
as those for the Edelstein effect.

\section{Intuitive picture for generation of phonon angular momenta by the temperature gradient}
\begin{figure}
\includegraphics[width = 8cm]{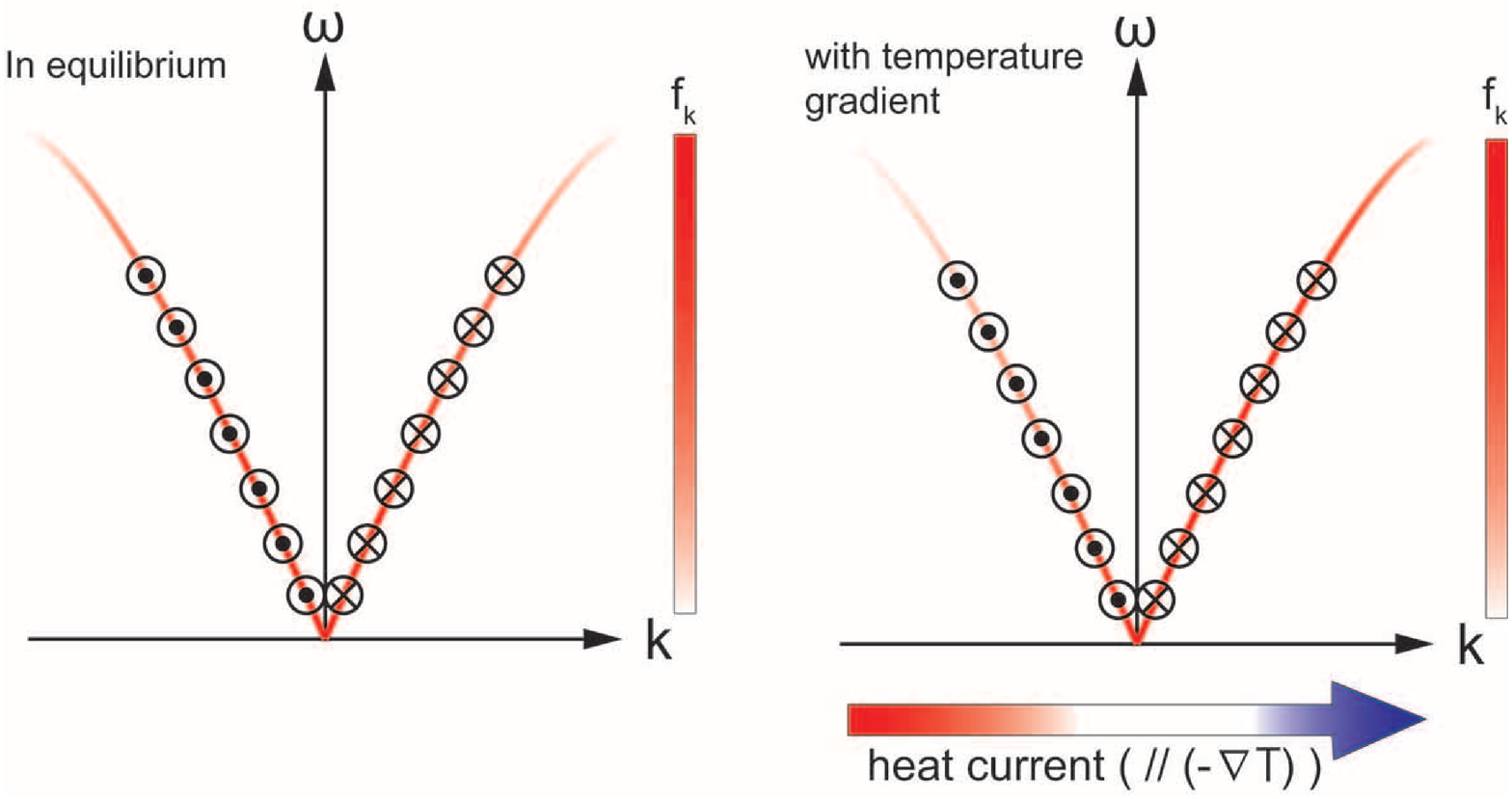}
\caption{\label{schematic-fig} Schematic figures of the phonon dispersion in equilibrium (left panel) and that with temperature gradient (right panel). The colormap shows the distribution of phonons following the distribution function $f_{\bm{k}}$. The circles with filled circles and crosses show the directions of phonon angular momentum, which are taken to be perpendicular to $k$}
\end{figure}

In FIG.~\ref{schematic-fig}, we show schematic figures for the direction of phonon angular momenta and distribution of each phonon mode with and without temperature gradient. The phonon distribution follows from the standard Boltzmann transport theory.
For simplicity, we only show the phonon band structure along the temperature gradient. In equilibrium, the heat current does not flow, because the phonon modes at $k(>0)$ and those at $-k(<0)$ are equally distributed. When the temperature gradient is in real space, the heat current flow, which means that the distribution of phonons at $k(>0)$ is larger than that at $-k(<0)$ as shown in the right panel of Fig.~\ref{schematic-fig}. Therefore, in the presence of the temperature gradient, the phonon angular momentum at $k$ and $-k$ no longer cancel, and nonzero phonon angular momentum is generated.
\section{Detalis of first-princeple calculations}
we describe the details of our first-principle calculation. The phonon properties in wurtzite GaN, Se, and Te crystals are calculated by using the density functional perturbation theory (DFPT)~\cite{RevModPhys.73.515} implemented in the Quantum-Espresso package~\cite{QE}.
We use the ultrasoft pseudopotentials generated by Garrity-Bennett-Rabe-Vanderbilt (GBRV) method~\cite{GBRV} for Ga, N, Se, and Te atoms.
The local-density approximation is adopted for the exchange-correlation functional
in GaN, and the generalized gradient approximation is adopted in Se and Te.
The expansion of the plane-wave set is restricted by a kinetic energy cutoff of $80$ Ry for GaN and $30$ Ry for Te and Se, respectively.
The optimized lattice constants in GaN are $a=3.155$~{\rm \AA} and $c= 5.143$~{\rm \AA}. 
The optimized intra-atomic/inter-atomic bond-length is $r=2.373$~{\rm \AA} and $R=3.440$~{\rm \AA} in Se and $r=2.834$~{\rm \AA} and $R=3.491$~{\rm \AA} in Te, respectively.
A $12\times12\times12~(6\times 6\times6)$ Monkhorst-Pack grid ~\cite{PhysRevB.13.5188} without an offset is used for self-consistent electronic structure calculations (phonon calculations) for GaN, and a $12\times12\times8~(8\times 8\times 8)$ grid is used for Se and Te. The nonanalytic term of the macroscopic electric field induced by optical phonon is also included in wurtzite GaN calculation.

\section{Valence force filed model of Keating for calculations for phonon angular momentum}
In this section we describe the details of the valence force field model of Keating, used for calculations of phonon angular momentum of GaN~\cite{Parameter01}. For simplicity, in this model we take the inter-atomic forces up to the next nearest neighbors.
The elastic energy of the valence force field model  is represented as 
\begin{equation}
\begin{split}
U_{i} &= \frac{3\alpha}{16r_{0}^{2}}\sum_{j=1}^{3} (\bm{r}_{ij}^2 - r_{0}^{2})^{2} + \frac{3\alpha^{\prime}}{16r_{0}^{\prime 2}}(\bm{r}_{i4}^{2} - r_{0}^{\prime 2})^{2}
\\
&+ \frac{3\beta}{8r_{0}^{2}} \sum_{j=1}^{3}\sum_{k>j}^{3}(\bm{r}_{ij}\cdot \bm{r}_{ik} - r_{0}^{2}\cos\theta_{0})^{2} 
\\
&+ \frac{3\beta^{\prime}}{8r_{0}^{\prime 2}}\sum_{k=1}^{3} (\bm{r}_{i4}\cdot\bm{r}_{ik} - r_{0}^{2}\cos\theta_{0}^{\prime})^2
\end{split}
\end{equation}
where $r_0$ is the equilibrium length between the Ga atom and the neighboring N atom, $r_{0}^{\prime}$ is the equilibrium length along $c$-axis between the Ga atom and the N atom, and $\theta_0$ and $\theta^{\prime}_0$ are bond angles shown in Fig.~\ref{GaNSup}(a).
is the equilibrium angle between two bonds, and $\theta_{0}^{\prime}$ is the equilibrium angle between a bond and the bond along the $c$-axis. The lattice constants are $a = 3.189 ~{\rm \AA}, c = 5.185 ~{\rm \AA}$ and the lattice parameter is $u = 0.3768$, which gives an equilibrium value of $r^{\prime}_0/c$. 
The four parameters $\alpha$, $\beta$, $\alpha^{\prime}$, and $\beta^{\prime}$ are Keating parameters, with $\alpha,\alpha^{\prime}$ being bond-stretching constants and $\beta,\beta^{\prime}$ being bond-bending constants. We 
adopt the Keating paramters as $\alpha = \alpha^{\prime} = 88.35 ~\rm{N/m}$ and $\beta =\beta^{\prime} = 20.92 ~\rm{N/m}$.  We calculate the inter-atomic force constants 
from this Keating potential, and calculate the dynamical matrix $D$. 
Here, for simplicity of the calculation we ignore the effect of polarization in wurtzite GaN in this model.

\section{Phonon angular momentum of Gallium Nitride by the valence force field model of Keating calculation}
\begin{figure*} 
\includegraphics[width=15cm]{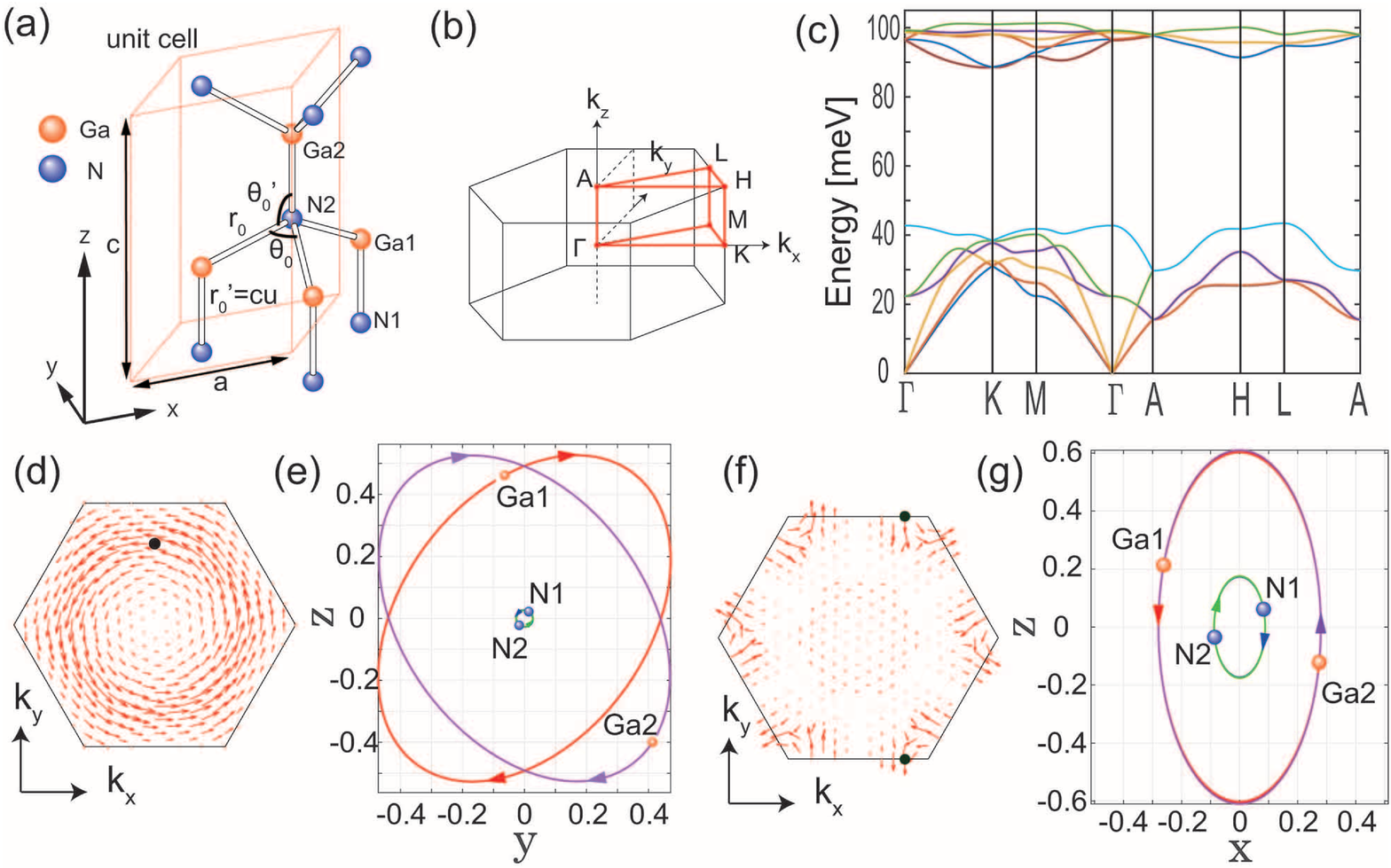}
\caption{\label{GaNSup} Crystal structure of GaN, and phonon angular momentum of GaN within the
valence force field model. 
(a) Crystal structure of wurtzite GaN.
(b) First Brillouin zone of GaN.
(c) Numerical results of phonon dispersion of wurtzite GaN using the valence force field model.  
(d) Distribution of the angular momentum of phonons $l_{\sigma}(\bm{k})$ of the sixth lowest band on the plane $k_z=0$. 
(e) Trajectories of the four atoms in the unit cell for the phonon mode of the sixth band at $\frac{{\bf k}a}{2\pi}=(0,0.3849,0)$, 
which is indicated as a black dot in (d).
(f) Distribution of the angular momentum of phonons $l_{\sigma}(\bm{k})$ of the third lowest band on the plane $k_z=0$. 
(g) Trajectories of the four atoms in the unit cell for the phonon mode of the third band at $\frac{{\bf k}a}{2\pi}=(0.2222,\frac{1}{\sqrt{3}},0)$,
which is indicated as a black dot in (f). 
(e) and (g) represent the normalized polarization vector $\varepsilon_{\sigma}({\bm k})$ with $\varepsilon^{\dagger}\varepsilon = 1$, and their axes $(x,y,z)$ are shown in a dimensionless unit.
}
\end{figure*}

 In this section we show the distribution of phonon angular momentum of GaN by the valence force field model of Keating. The crystal structure of wurtzite GaN is shown in Fig.~\ref{GaNSup}(a), and the phonon dispersion is obtained as Fig.~\ref{GaNSup}(c) with the Brillouin zone in Fig~\ref{GaNSup}(b).  In this model the inter-atomic force is considered up to next nearest neighbors and the effect of polarization is ignored for simplicity. Nevertheless, despite its simplicity, this model well describes the nature of phonon angular momentum. The phonon bands below the frequency gap are in good agreement with first-principle calculation, but the phonon bands above the frequency gap are not. Nevertheless, this
model is 
useful for extracting the distribution of phonon angular momentum
in a semi-quantitative manner.  In Figs.~\ref{GaNSup}(d) and (f), we show the distributions of the angular momentum of phonons $l_{\sigma}(\bm{k})$of the sixth and third lowest bands on the plane $k_z = 0$. As illustrative examples, Figs.~\ref{GaNSup}(e) and (g) show the trajectories of the four atoms in the unit cell for particular choices of the mode $\sigma$ and the wavenumber $\bf{k}$ shown as black dots in Figs.~\ref{GaNSup}(d) and (f), 
respectively.
Thus, the model calculations show similar results with our first-principle calculations.



\section{Phonon angular momentm of Tellurium and Selenium}
\begin{figure*} 
\includegraphics[width=15cm]{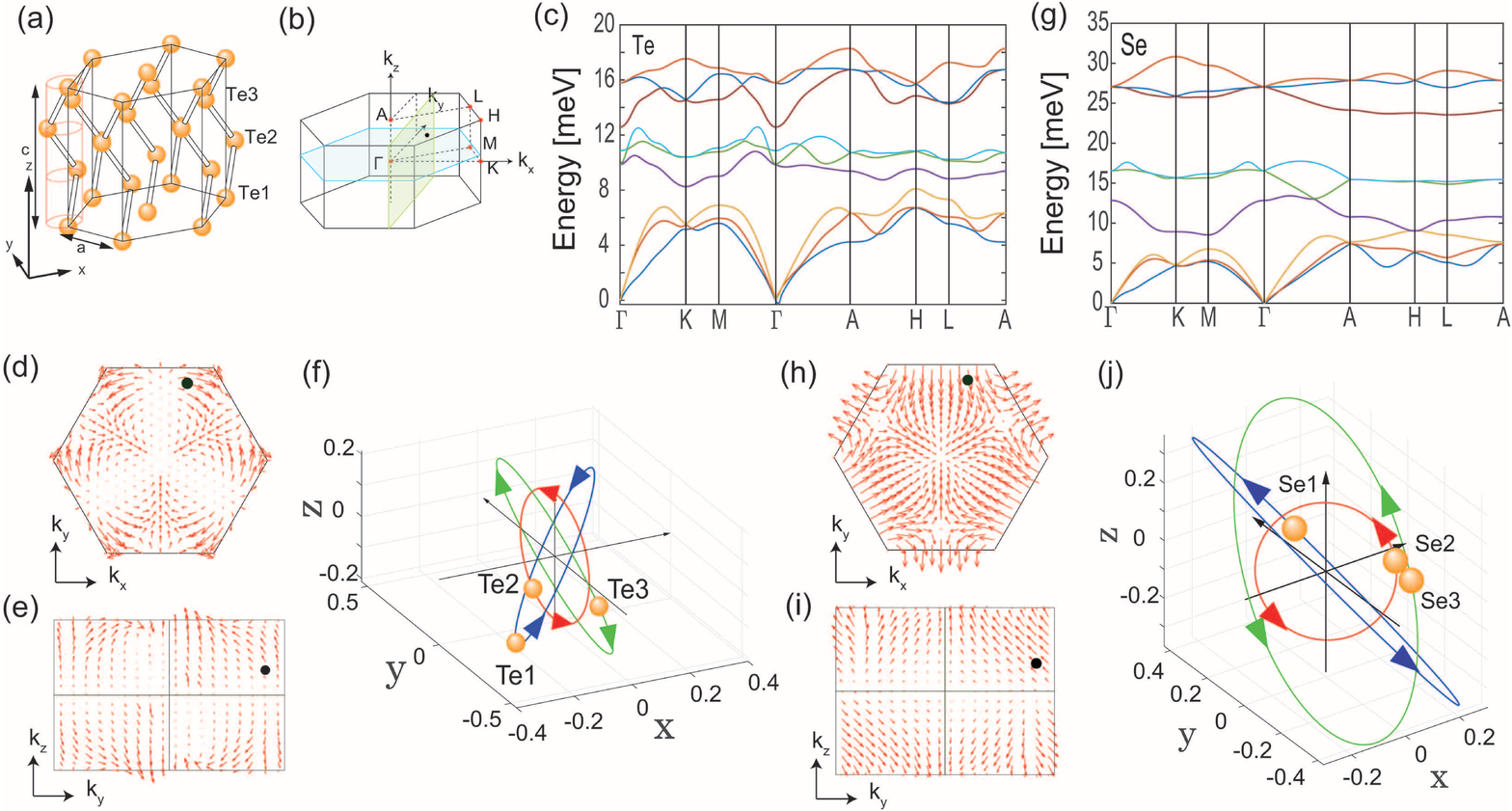}
\caption{\label{TeSeSup} Phonon angular momentum of Te and Se in the first-principle calculation. 
(a) Crystal structure of Te and Se. 
(b) First Brillouin zone of Te and Se. 
(c) Numerical results of phonon dispersion of Te.
(d),(e) Distribution of the angular momentum of phonons $l_{\sigma}(\bm{k})$ of the third lowest band. 
(d) and (e) show the results on the plane $\frac{k_za}{2\pi}= 0.1254$ and on the plane $\frac{k_xa}{2\pi} = \frac{1}{6}$, respectively.  These planes correspond to two cross sections in (b). 
(f) Trajectories of the four atoms in the unit cell for the phonon of the fourth lowest band at $\frac{{\bf k}a}{2\pi}=(\frac{1}{6},0.4811,0.1254)$, 
which is indicated as a black dot in (d) and (e).
(g) Numerical results of phonon dispersion of Se.
(h),(i) Distribution of the angular momentum of phonons $l_{\sigma}(\bm{k})$ of the third lowest band. 
(h) and (i) show the results on the plane $\frac{k_za}{2\pi}=0.1471$ and on the plane $\frac{k_xa}{2\pi}=\frac{1}{6}$, respectively.  These planes correspond to two cross sections in (b). 
(j). Trajectories of the four atoms in the unit cell for the phonon of the third lowest band at $\frac{{\bf k}a}{2\pi}=(\frac{1}{6},0.4811,0.1471)$, 
which is indicated as a black dot in (h) and (i). 
(f) and (j) represent the normalized polarization vector $\varepsilon_{\sigma}({\bm k})$ with $\varepsilon^{\dagger}\varepsilon = 1$, and their axes $(x,y,z)$ are shown in a dimensionless unit.
}
\end{figure*}

In this section, we show the distribution of phonon angular momentum of Te and Se by the first-principle calculation. Te and Se have a helical crystal structure, as shown in Fig.~\ref{TeSeSup}(a). The numerical results of the phonon dispersions of Te and Se are shown in Fig.~\ref{TeSeSup}(c) and (g), respectively, with the Brillouin zone in Fig.~\ref{TeSeSup}(b). These dispersions are qualitatively similar. The distributions of phonon angular momentum $l_{\sigma}(\bm{k})$ within the two planes in the Brillouin zone (Fig.~\ref{TeSeSup}(b)) are shown in Figs.~\ref{TeSeSup}(d) and (e) for the third lowest band of Te and in Fig.~\ref{TeSeSup}(h) and (i) for the third lowest band of Se. Figures~\ref{TeSeSup}(f) and (j) represent the trajectories of the three atoms in the unit cell for Te and Se, respectively. The overall features of the distributions for Te and Se are qualitatively similar, while the sizes of the phonon angular momentum are somewhat different. In addition, the trajectories of the three atoms in the unit cell for Te and Se at the same wavenumber $\bm{k}$ are quite different, as seen in Figs.~\ref{TeSeSup}(f) and (j). 

\section{numerical results of induced phonon angular momentum and a rigid-body rotation}
In this section, we show numerical results of the induced phonon angular momentum and generated rigid-body rotation.  When we assume the phonon relaxation time of GaN is $\tau \sim 10~{\rm ps}$~\cite{parameter02} with the temperature $T=300~{\rm K}$, the numerical results are shown in Tab.~\ref{T1}. From Eq.~(\ref{GaN}), nonzero components of response function are $\alpha_{xy}$ and $\alpha_{yx}(=-\alpha_{xy})$. In Tab.~\ref{T1}, the response function $\alpha_{xy}$ below the frequency gap means  the sum from the lowest band to the sixth lowest band in Fig.~\ref{GaNSup}, and the response function $\alpha_{xy}$ above the frequency gap means the sum from the seventh lowest band to the highest band. If we take the temperature difference to be $\Delta T = 10~{\rm K}$, 
the angular velocity of a rigid-body rotation by first-principle calculation is $3.4\times10^{-8}~\rm{s^{-1}}$ for $L = 100~{\rm \mu m}$ and 
$3.4\times10^{-2}~{\rm s^{-1}}$ for $L = 1~{\rm \mu m}$. This rigid-body rotation is expected to be experimentally observable. 

Next, we show the numerical results of Te and Se for $T=300~{\rm K}$ in Tab.~\ref{T2}. We assume the phonon relaxation times along and perpendicular to c-axis of Te are $\tau_{\parallel}\sim10~{\rm ps}$ and $\tau_{\perp}\sim1~{\rm ps}$, and those of Se are $\tau_{\parallel}\sim10~{\rm ps}$ and $\tau_{\perp}\sim10~{\rm ps}$, respectively~\cite{Coorper,Springer,doi:10.1063/1.4938203}.  From Eq.~(\ref{TeSe}), the nonzero components of response function are $\alpha_{xx} = \alpha_{yy}$ and $\alpha_{zz}$. Therefore, when the temperature gradient along the 
$z$-axis is nonzero, the phonon angular momentum along the $z$ axis is induced, and when temperature gradient along the $y$-axis is nonzero, the phonon angular momentum is 
along the $y$ axis.  In this setup, the induced phonon angular momentum and a rigid-body rotation are expected to be experimentally observable, and the induced phonon angular momentum in Se is larger than that of Te.
\begin{table*}
\caption{\label{T1}{Numerical results of induced phonon angular momentum and a rigid-body rotation of GaN } }
\begin{tabular}{lrr} \hline\hline
 & valence force field model & first-principle \\ \hline
$a~[{\rm \AA}]$ & 3.189 & 3.155 \\
$c~[{\rm \AA}]$& 5.185 & 5.143 \\
$\rho~[{\rm kg/m^3}]$ & $6.089\times10^{3}$ & $6.276\times10^{3}$ \\ 
$\alpha_{xy}$ below the frequency gap $~[\rm Jsm^{-2}K^{-1}]$& $-7.4\times10^{-8}\times \tau$ & $-8.9\times10^{-8}\times \tau$ \\
$\alpha_{xy}$ above the frequency gap $~[\rm Jsm^{-2}K^{-1}]$ & $1.8\times10^{-7}\times \tau$ & $-2.7\times10^{-7}\times \tau$ \\
total $\alpha_{xy}~[\rm Jsm^{-2}K^{-1}]$ & $1.1\times10^{-7} \times \tau$ & $-3.6\times10^{-7}\times \tau $ \\
$\omega_{\rm rigid-body}~[{\rm s^{-1}}]$&$ -1.1\times10^{-21}\times \Delta T/L^3$&$3.4\times10^{-21}\times\Delta T/L^3$ \\ \hline\hline
\end{tabular} 
\end{table*}

\begin{table*}
\caption{\label{T2}{Numerical results of induced phonon angular momentum and a rigid-body rotation of Te and Se} }
\begin{tabular}{lrr} \hline\hline
 & Te & Se \\ \hline
$a~[{\rm \AA}]$ & 4.456 & 4.371 \\
$c~[{\rm \AA}]$& 5.921 & 4.954 \\
$\rho~[{\rm kg/m^3}]$ & $6.245\times10^{3}$ & $4.800\times10^{3}$ \\ 
$\tau_{\parallel}~[{\rm ps}] $& $\sim 10$ & $\sim 10$ \\
$\tau_{\perp}~[{\rm ps}] $ & $\sim 1$ & $\sim 10$ \\
$\alpha_{xx}~[\rm Jsm^{-2}K^{-1}]$ & $-3.0\times10^{-7}\times \tau_{\perp}$ & $-7.1\times10^{-7}\times \tau_{\perp}$ \\
$\alpha_{zz}~[\rm Jsm^{-2}K^{-1}]$ & $4.8\times10^{-7}\times \tau_{\parallel}$ & $-9.7\times10^{-6}\times \tau_{\parallel}$ \\
$\omega_{\rm rigid-body}$ perpendicular to c-axis $~[{\rm s^{-1}}]$&$ 2.9 \times10^{-22}\times \Delta T/L^3$&$8.9\times10^{-21}\times\Delta T/L^3$ \\
$\omega_{\rm rigid-body}$ along c-axis$~[{\rm s^{-1}}]$&$ -4.6\times10^{-21}\times \Delta T/L^3$&$1.2\times10^{-19}\times\Delta T/L^3$ \\ \hline\hline
\end{tabular} 
\end{table*}

\section{Conservation of the sum of the phonon angular momentum and the angular momentum of the rigid-body rotation of the crystal}
In the main text, we propose an experiment  of measuring the phonon angular momentum as a rigid-body rotation of the whole crystal. It is based on the conservation of the sum 
of the phonon angular momentum $J^{\rm{ph}}$ and the angular momentum  $J^{\rm{rigid-body}}$ 
of the rigid-body rotation  of the crystal. 
In this section, we discuss how it is conserved in the present case. 
 Here, let $\bm{R}_j$ denote the equilibrium position of the $j$-th atom, and and $\bm{u}_j$ denote 
its deviation from the equilibrium position. We define the mass of the $j$-th atom $M_j$. 
Then the conservation of angular momentum is written 
as $\sum_jM_j(\dot{\bm{R}}_j+\dot{\bm{u}}_j)\times(\bm{R}_j+\bm{u}_j)=0$. We average this quantity over
a long period, which is much longer than the typical timescale of the phonon motions. Then, the typical time scale of $\bm{R}$ is $10^2$-$10^8$sec as we estimate in the main text, and it is much longer than the time scale of the phonon motion. Therefore, the time-averages of the cross terms 
 $\sum_jM_j\dot{\bm{R}}_j\times\bm{u}_j$, and  $\sum_jM_j\dot{\bm{u}}_j\times\bm{R}_j=0$
vanish and are neglected.
Then  after time-averaging, the sum of 
 $J^{\rm{ph}}=\sum_jM_j\dot{\bm{u}}_j\times\bm{u}_j$, and  
 $J^{\rm{rigid-body}}=\sum_jM_j\dot{\bm{R}}_j\times\bm{R}_j$ is zero.

%